\documentclass[3p,preprint,11pt]{elsarticle}

\usepackage{amssymb,amsmath,amsthm,bm,url}

\biboptions{sort&compress}

\usepackage{graphicx}
\graphicspath{{figures-pdf1/}{figures-pdf/}}

\usepackage{color}
\definecolor{dgreen}{rgb}{0,.6,0}

\newlength\imagewidth
\newlength\figwidth
\usepackage[mathlines,displaymath]{lineno}

\usepackage[normalem]{ulem}

\begin{document}

\begin{frontmatter}

\title{An optical image encryption scheme based on depth-conversion integral imaging and chaotic maps}

\author[kr-ysu]{Xiaowei Li}

\author[cn-xtu]{Chengqing Li}
\author[kr-pku]{Seok-Tae Kim}
\author[kr-ysu]{In-Kwon Lee\corref{corr}}
\ead{iklee@yonsei.ac.kr}
\cortext[corr]{Corresponding author.}

\address[kr-ysu]{Department of Computer Science, Yonsei University, Seoul, Republic of Korea}

\address[cn-xtu]{College of Information Engineering, Xiangtan University, Xiangtan 411105, Hunan, China}

\address[kr-pku]{Department of Information and Communications, Pukyong National University, Busan, Republic of Korea}

\begin{abstract}
Integral imaging-based cryptographic algorithms provides a new way to design secure and robust image encryption schemes. In this paper, we introduce a
performance-enhanced image encryption schemes based on depth-conversion integral imaging and chaotic maps, aiming to meet the requirements of
secure image transmission. First, the input image is decomposed into an elemental image array (EIA) by utilizing a pinhole array. Then,
the obtained image are encrypted by combining the use of cellular automata and chaotic logistic maps. In the image reconstruction process, the conventional computational integral imaging reconstruction (CIIR) technique is a pixel-superposition technique; the resolution of the reconstructed image is dramatically degraded due to the large magnification in the superposition process as the pickup distance increases. The smart mapping technique is introduced to improve the problem of CIIR. A novel property of the proposed scheme is its depth-conversion ability, which converts original elemental images recorded at long distance to ones recorded near the pinhole array and consequently reduce the magnification factor. The results of numerical simulations demonstrate the effectiveness and security of this proposed scheme.
\end{abstract}


\end{frontmatter}

\section{Introduction}
\label{sec:intro}

With the development of network and the increasing requirement for image transmission, image security becomes a very paramount issue in communication science.
All kinds of security problems about image data emerged, such as unauthorized access \cite{YaobinMao:CSF2004,Lee:mosaic:ITCSVT14} and forge detection \cite{Li:Copy-move:TIFS15}. Encrypting image as text is the most direct way to protect image. Due to strong correlation existing in image data, the conventional text encryption techniques are not appropriate for image encryption. In the past two decades, a large number of schemes have been proposed for image encryption by utilizing all kinds of optical techniques \cite{REFREGIER:OL:1995,Hennelly:OL:2003,Abuturab:AO:2013}, which own distinct advantages of processing two-dimensional (2D) image data with parallel conduction mode. Among them, the most widely transform methods are fractional Fourier transform (FRT), extended FRT, gyrator transform and Fresnel transform. As some properties of chaos are very similar to that of a secure encryption scheme, such as high sensitivity to initial values and system parameters,
chaotic maps were used to design image encryption schemes \cite{YaobinMao:CSF2004,Ye:Scramble:PRL10,Dogaru:ITCSVT:2010,Bhatnagar:ITSC:2014,Zhou:ITC:2013}.
Meanwhile, much works on analyzing security level of existing encryption schemes were presented \cite{Li:AttackingISWBE2006,Zhou:ITC:2013,Anstett:ITCSIP:2006,AlirezaL:3Dcrypt:TIFS15,Lee:mosaic:ITCSVT14,
Li:AttackingBSSE2006,LiLi:defects:IVC09,Li:permutation:SP11}.

In recent years, the integral imaging technique has received intensive attention in the study of image encryption \cite{Piao:integral:OLE2009,Lixw:neural:OC2014,Lixw:integral:Optik2014}. It divides an input image into many elemental images by utilizing a lenslet array or a pinhole array. Each elemental image has nearly the full attributes of the input image. Thereby, encrypting the elemental images can greatly improve the robustness and security in image transmission. The conventional integral imaging system is composed of two parts: pickup part and reconstruction part. In integral imaging pickup and reconstruction processes, an optical lenslet array is utilized in both processes to capture and reconstruct the input image. In the pickup process, the light rays coming from the input image pass through the lenslet array and recording a set of inversely small images by using a pickup device. The recorded small images array is referred to as elemental image array (EIA). In the image reconstruction process, light rays of the elemental images by back-propagating through the lenslet array, the input image is reconstructed when it was originally picked up. In the case of utilizing the optical devices, the quality of the reconstructed image is poor due to the light diffraction and limitation of the optical devices.

To cope with this problem, some computational reconstruction techniques were proposed \cite{Shin:integral:OE2008,Lee:reconstruction:OC2008,Hwang:reconstruction:AO2008}. Among them, the most commonly used method is the computational integral imaging reconstruction (CIIR) algorithm, which can digitally reconstruct image by using the recorded EIA. Each of the recorded elemental images are inversely projected onto an output plane and inversely magnified according to a magnification factor. This method can improve the image resolution because of diffraction freedom and absence of optical device. The CIIR technique, however, remains some serious problems, such as the image resolution seriously degrades as the increase of the reconstruction distance. Because CIIR is pixel reversely magnified reconstruction technique, the pixel-overlapping area will be larger as the magnification factor increase. The interference of adjacent pixels will degrade the quality of the reconstructed image. Many researchers focused on how to improve the quality of the reconstructed image in integral imaging systems. However, many techniques are still utilized to the CIIR technique \cite{Hong:integral:OE2004}, the resolution degradation of the input image over the far distance from the lenslet array is not greatly improved.

In this paper, we present a new image security system based on the smart mapping technique to improve the quality of the reconstructed image.
Smart mapping is a depth-conversion process in which the original elemental images recorded at long distance can be converted to ones recorded near the pinhole array. This algorithm can provide improved quality of the reconstructed image because of reducing the interference problem by decreasing the magnification factor. Meanwhile, the robustness of encrypted image can be greatly improved owing to the property of data redundancy of 2D EIA. To show the usefulness of the proposed scheme, some experiments are carried out for several typical test images and its results are presented.

The rest of this paper is organized as follows. Section~\ref{sec:review} review previous related work on image encryption.
Then, Sec.~\ref{sec:analysis} gives some theoretical analysis. The proposed image encryption scheme is proposed in Sec.~\ref{sec:scheme} and
detailed experimental results are presented in Sec.~\ref{sec:experiment}. Finally, the last section concludes the paper.

\section{Previous Work}
\label{sec:review}

In \cite{Shin:integral:OE2008}, Hwang et al. proposed an image copyright protection scheme based on computational integral imaging (CII) technique, in which the 3D watermark information is embedded into the DWT domain. The 3D watermark plane image can be reconstructed by using CIIR technique. This scheme provides high robustness because of the data redundancy property of the elemental images. However, in the watermark extraction process, CIIR is a pixel superposition reconstruction method. The resolution of the reconstructed 3D plane image is dramatically degraded as the magnification factor increases.
In \cite{Piao:integral:OLE2009}, Piao et al. developed an image encryption scheme based on integral imaging and pixel scrambling technique. In the scheme, the input image is firstly recorded by the CII technique and the obtained elemental images are scrambled by pixel scrambling technique. In the reconstruction process, the image is reconstructed by the CIIR technique. This method provides high security and robust encoding system. However, it causes the problem of resolution degrades caused by CIIR.

In our previous works proposed in \cite{Lixw:neural:OC2014}, we presented a multiple-image encryption scheme based on CII and back-propagation (BP) neural network. In the multiple-image encryption part, a CII pickup technique is employed to record the multiple-image simultaneously to form an EIA. The EIA is then encrypted by using the maximum length cellular automata and random phase encoding algorithm. To improve the problem of the low-resolution caused by CIIR, the BP neural network is utilized in the scheme. Numerical simulations have been performed to demonstrate the performance of the method. However, the reference EIAs is needed in the process of the BP network training. When decrypting the multiple-image, the sender record all reference EIAs correctly and send them to the receiver. The encryption process is a little complicate and the network load will be huge. In \cite{Lixw:integral:Optik2014}, we proposed 3D image encryption scheme based on the computer-generated integral imaging (CGII) and cellular automata transform, in which a 2D EIA is digitally recorded by lights coming from the 3D image through a pinhole array according to the race tracing theory. Cellular automata (CA) transform is used to encrypt the 2D EIA. This scheme can provide high level of security because of the hologram-like property of 2D EIA and huge key space provided by cellular automata.

In \cite{Baptista:Crypt:PLA1998}, Maptista presented a message encryption scheme based on chaotic logistic map, where each alphabet
is encrypted into an iteration number of the map falling in given interval of the phase space. In \cite{Pareek:IVC2006}, Pareek et al. proposed an image encryption scheme using the chaotic logistic map, where an external secret key with 80-bit and two chaotic logistic maps were used. To make the encryption system more robust against attacks, in this method, the secret key needs to be modified after encrypting each block the image. However, the scheme were found to be insecure against differential attack \cite{LiLi:defects:IVC09}.

\section{Theoretical analysis}
\label{sec:analysis}

\subsection{Review of conventional CGII}

\begin{figure}[!htb]
\centering
\begin{minipage}[t]{\figwidth}
\centering
\includegraphics[width=\imagewidth]{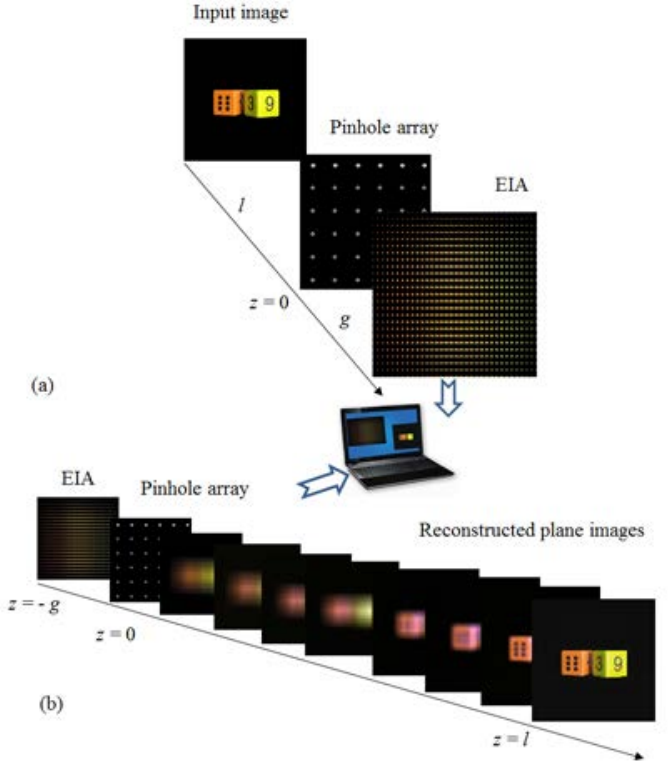}
\end{minipage}
\caption{Schematic diagram of CGII system.}
\label{fig:diagram}
\end{figure}

A CGII system uses a set of elemental images generated by computer graphics, instead of the pickup process using the optical lens array \cite{Hong:integral:OE2004,Hwang:integral:OC2007,Shin:integral:OE2008,Hwang:reconstruction:AO2008}. As illustrated in Fig.~\ref{fig:diagram}, conventional CGII system are composed of pickup and reconstruction phases. In the pickup process shown in Fig.~\ref{fig:diagram}a), rays emitted from the input image are computationally recorded by a virtual pinhole array. In the computer reconstruction process, the reconstruction is the inverse process of the pickup process; the elemental images are projected on the image plane inversely through each virtual pinhole. The EIA can be inversely magnified according to the magnification factor of $M =l/g$, where variable $l$ represents the distance between the virtual pinhole array and the input image, and varible $g$ denotes the gap between the virtual pinhole array and the EIA. The reconstructed images can be arbitrarily displayed along the output plane. Figure~\ref{fig:diagram}b) dynamically displays the reconstructed plane images along the output plane $z=l$. The plane image can be clearly reconstructed at the location where it was originally picked up.
The CGII pickup and reconstruction processes can be implemented according to equation
\begin{equation}
E_{ij}(x,y)=O\left(-\frac{xg}{l}+i\sigma, -\frac{yg}{l}+j\sigma\right),
\end{equation}
where $E_{ij}(x,y)$ represents the $(i, j)$-th elemental image, $x$ and $y$ denote the coordinates of the $(i, j)$-th pinhole,
$\sigma$ is the pitch of a pinhole, $l$ is the distance between the virtual pinhole array and the input plane image, and $g$ is the gap between the EIA and virtual pinhole array. Meanwhile, the computer reconstruction process can be calculated by
\begin{equation*}
O_r(x, y)=\sum_{i=0}^{p-1}\sum_{j=0}^{q-1}O\left(-\frac{xl}{g}+i\sigma, -\frac{yl}{g}+j\sigma\right),
\end{equation*}
where $O_r(x, y)$ represents the reconstructed plane image and $p\times q$ represents the number of elemental images.

\subsection{Limitation of CGII}

\begin{figure}
\center
\includegraphics[width=\imagewidth]{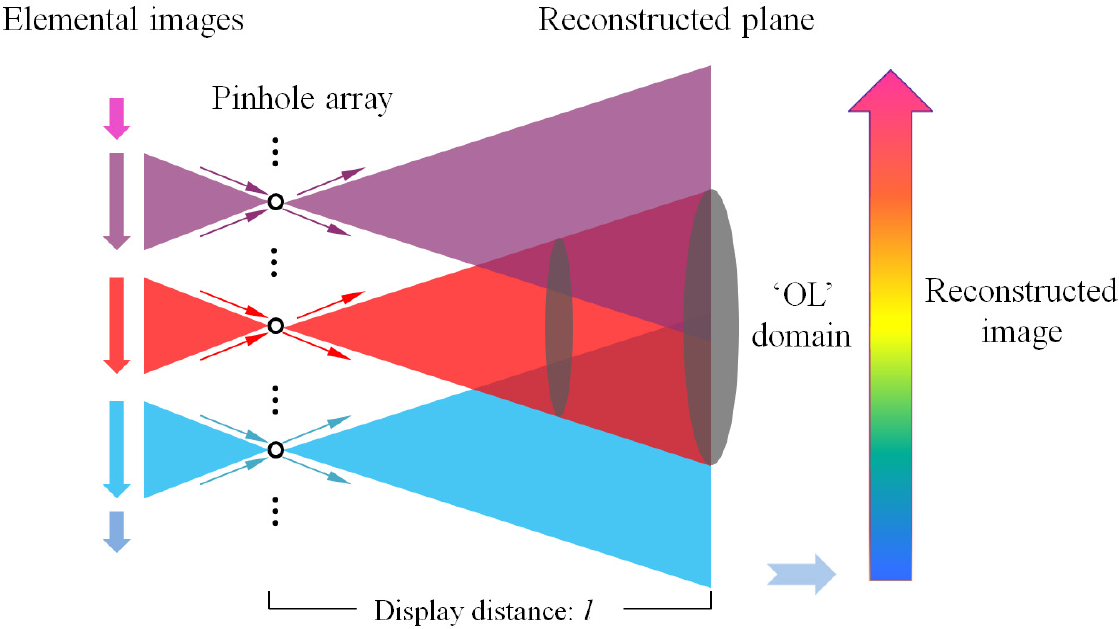}
\caption{Principle of pixel mapping for conventional CIIR algorithm. Here, 'OL' domain denotes overlapping domain}
\label{fig:mapping}
\end{figure}

Even though CGII was considered as a promising technique for the image security, so far it still have some weak points. One of them is the limitation of the effect pickup distance of the input image. Figure~\ref{fig:mapping} illustrates the computer reconstruction process of image on the output plane. Each elemental images is inversely magnified mapped through each virtual pinhole according to the magnification factor $M$. As the magnification factor $M$ is proportional to the reconstruction distance $l$, the greater the reconstruction distance $l$ is, the more the reconstructed image is dispersedly spread in intensity.
This point can be observed in Fig.~\ref{fig:mapping}. Meanwhile, the interference strength of adjacent pixels becomes larger with the increase of the pickup distance $l$.

\subsection{Analysis of depth-conversion algorithm using pixel to pixel smart mapping}

\begin{figure}
\center
\includegraphics[width=1.1\imagewidth]{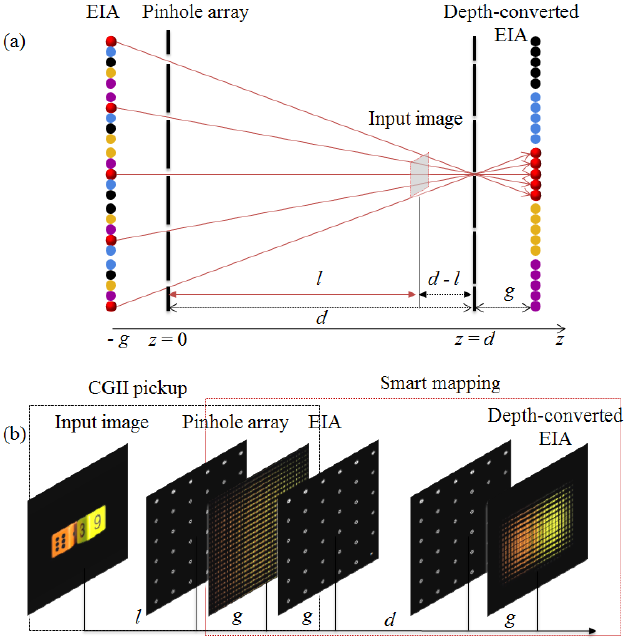}
\caption{Schematic diagram of CGII system: (a) pickup stage and (b) reconstruction stage.}
\label{fig:EIA_Diagram}
\end{figure}

Figure~\ref{fig:EIA_Diagram} shows the principle of the depth-converted EIA mapping method. The EIA depth-conversion method also is called smart mapping algorithm \cite{Martinez-Corral:OE:2005,Shin:AO:2008}. Smart mapping method is a computational depth-conversion process in which the original picked-up elemental images are convertible to ones recorded at the different pickup distances. As shown in Fig.~\ref{fig:EIA_Diagram}(a), the elemental images recorded from one input image at the long distance ($z=l$), which can be converted to ones recorded near the pinhole array ($z=d-1$). For smart mapping process, a pinhole array should be located at the distance of and another pinhole array located at the fixed distance of $z=0$ as shown in Fig.~\ref{fig:EIA_Diagram}(a). The condition for the fixed distance $d$ can be easily derived from geometrical mathematics, $d=m\times g$, where $m$ denotes the pixel numbers of each elemental image and $g$ is the gap between the pinhole array and EIA. The depth-converted EIA can be synthesized in the pickup distance of $z=d+g$ and the smart mapping algorithm for the one-dimensional case is formulated as
\begin{equation*}
E^d_{i,j}=E_{pq}
\end{equation*}
and
\begin{equation*}
q=(M-1)-j,
\end{equation*}

\begin{equation*}
p=\begin{cases}
  i+\frac{M}{2}-j    & \mbox{if } M \mbox{ is even};\\
  i+\frac{M+1}{2}-j  & \mbox{otherwise},
\end{cases}
\end{equation*}
where $E^d$ and $E$ represent the depth-converted EIA and the original EIA, respectively. The subscript sets $(i, p)$ and $(j, q)$ correspond to the number of elemental images and number of intensity values, respectively. Figure~\ref{fig:EIA_Diagram}(b) shows the two-dimensional simulation model of the depth-converted EIA generation. After performing the smart mapping process, the depth-converted EIA can be obtained. Thereby, the far image can be reconstructed at a short distance by the pinhole array. In other words, the quality of the reconstructed image is greatly improved by reducing the interference of adjacent pixels as the decrease of the reconstructed distance.

\subsection{2D pseudo-random mask}

\begin{figure*}
\center
\includegraphics[width=2\imagewidth]{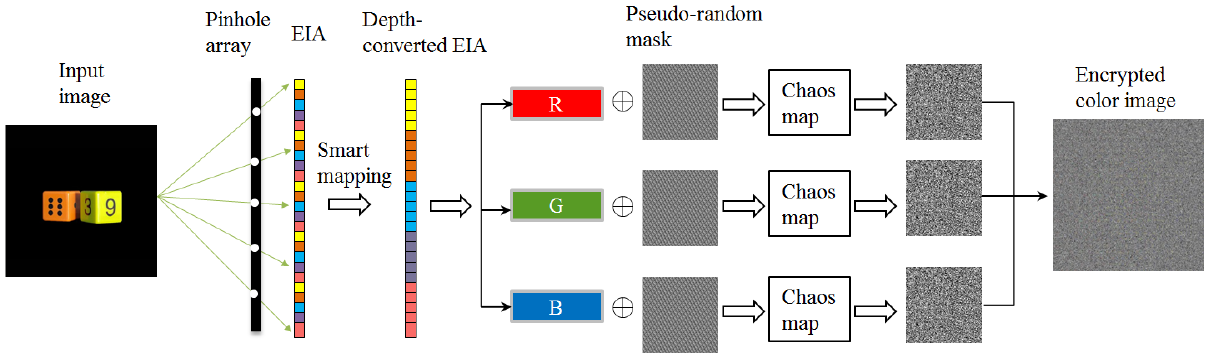}
\vspace{5pt}
\caption{Flow chart of the proposed encryption scheme.}
\label{fig:flowchart}
\end{figure*}

A high quality 2D pseudo-random mask can be generated by a maximum-length CA. In this paper, we adopt the 2D pseudo-random mask for image encryption for the first time with careful considerations for the confusion and diffusion and possible attacks. The evolution of the CA serves to confuse the plaintext into an unrecognizable form. This evolution also provides diffusion. Information stored in a cell is spread to the entire neighborhood in each evolution step \cite{WOLFRAM:AAM:1986}. Using a three-site CA, a 256-byte block, and a key of period 256, information in a cell spreads to every other cell in the CA at encryption time. The 2D pseudo-random mask generation method was investigated deep in our previous work \cite{Piao:integral:OLE2009}. The pseudo-random mask is used in this experiment as shown in Fig.~\ref{fig:flowchart}, which is generated by the combined use of the CA rule 90 and rule 150. Rule 90 and rule 150 can be defined as
\begin{equation*}
\alpha^{t+1}_i=\alpha^{t}_{i-1}\oplus \alpha^{t}_{i+1},
\end{equation*}
\begin{equation*}
\alpha^{t+1}_i=\alpha^{t}_{i-1}\oplus \alpha^{t}_{i}\oplus \alpha^{t}_{i+1},
\end{equation*}
where $\alpha^t_i$ represents the state of the node $i$ at time $t = k$.

\subsection{Pixels Scrambling by Chaotic maps}

In this scheme, we introduce one-dimensional chaotic map to encrypt the hiding EIA. Chaotic sequence is sensitive to initial value and unpredictability, so it is widely combined with traditional encryption \cite{Lang:OC:2010,Li:permutation:SP11}. It is one of the simplest nonlinear chaotic discrete systems that display chaotic behavior defined by equation
\begin{equation}
x_{i+1}=\rho_0 x_i(1-x_i),
\label{eq:logistic}
\end{equation}
where $x_0\in(0, 1)$ denotes an initial input value, $\rho_0$ $(0<\rho_0<4)$ is the system parameter and $x_i$ is the state value.
When $\rho_0\in(3.5699456, 4)$, the slight variations in the initial value yield dramatically different results over time. In other words, logistic map will operate in chaotic state. The sequence generated by Eq.~(\ref{eq:logistic}), which is used to encrypt the hiding EIA. With $x0$ as an initial value, which is also known as the seed for the chaotic map system, a non-periodic and non-converging sequence $\{x_i\}$ can be generated iteratively by using Eq.~(\ref{eq:logistic}). Assume that an image $O_{MN}$ of size $M\times N$ is scrambled by the chaos function, the scrambling process is briefly described as follows:
\begin{itemize}
\item  Generate the chaos sequence $S=\{x_i\}_{i=1}^{MN})$ with the given initial parameters ($x0$, $\rho_0$) by using Eq.~(\ref{eq:logistic}). Here, we give the $x_0 = 0.1775727$ and $\rho_0=3.5725212$.

\item Rearrange the sequence S in ascending or descending orders to form a new sequence $R$. The values of the elements in the sequence are not altered but the positions are changed. Here, we need to record the pixel permutation mapping index from the sequence $S$ to the sequence $R$. That means the $i-$th element in $R$ corresponds to the $i$-th element in $S$.

\item Convert the 2D image matrix into 1D sequence factor $V=\{a_i\}_{i=1}^{MN}$. $V$ is scrambled as $V'=\{a_i'\}_{i=1}^{MN}$.
Then $V'$ is reconverted into a new 2D matrix $O'_{MN}$, which is the final encoding image.
\end{itemize}

\section{Proposed image encryption scheme}
\label{sec:scheme}

\subsection{Encryption process}

In a standard RGB image, the color models are mathematical representations of a set of colors. R, G and B channels are equivalent and the same encryption method is parallel employed to each channel. The schematics of the proposed image encryption scheme is described in Fig.~\ref{fig:flowchart}. The image encrpytion process in this scheme can be divided into five steps.
\begin{itemize}
\item The input color image $O$ is divided into three channels and each component is equally computationally recorded by CGII in the form of an EIA,
where initial parameters of the CGII pickup system ($l = 69$ mm, $g = 3$ mm), input parameters of 2D chaotic logistic map ($x_0$, $\rho_0$), generation keys $(150 90 150 90 90 90 150 90)$ of the pseudo-random mask.

\item The recorded EIA is transformed into a depth-converted EIA by utilizing the smart mapping algorithm.

\item Each channel of the depth-converted color EIA is msked by utilizing a pseudo-random sequence:
\begin{equation*}
f_h(x, y)=\sum_{x=0}^{M-1}\sum_{y=0}^{N-1}\{E^d(x, y) \oplus M(x, y)\},
\end{equation*}
where $M\times N$ denotes the size of the input image, $E^d(x, y)$ and $M(x, y)$ represent the depth-converted EIA and the pseudo-random sequence.

\item The hidden depth-converted EIA is finally scrambled by the permutation sequence generated by the chaotic logistic map, Eq.~(\ref{eq:logistic}).

\item The three scrambled channels are combined to obtain the color encrypted image.
\end{itemize}

\subsection{Decryption process}

The image decryption is the inverse process of the image encryption. The image encryption process can be briefly illustrated as follows.
\begin{itemize}
\item The inverse process of chaotic logistic map is applied to the encoded image and the hidden depth-converted EIA is obtained;

\item The depth-converted EIA can be recovered by using the inverse pseudo-random masking algorithm as
\begin{equation*}
R^d(x, y)=\sum_{x=0}^{M-1}\sum_{y=0}^{N-1}\{E^*(x, y) \oplus \Psi(x, y)\},
\end{equation*}
where $E^*(x, y)$ denotes the masked depth-converted EIA.

\item By utilizing the computational reconstruction technique, the high quality image can be reconstructed from the recovered depth-converted EIA.
To further remove the blur noise of the reconstructed image caused by CIIR.
\end{itemize}

\section{Simulation results and discussion}
\label{sec:experiment}

\begin{figure}
\center
\includegraphics[width=2\imagewidth]{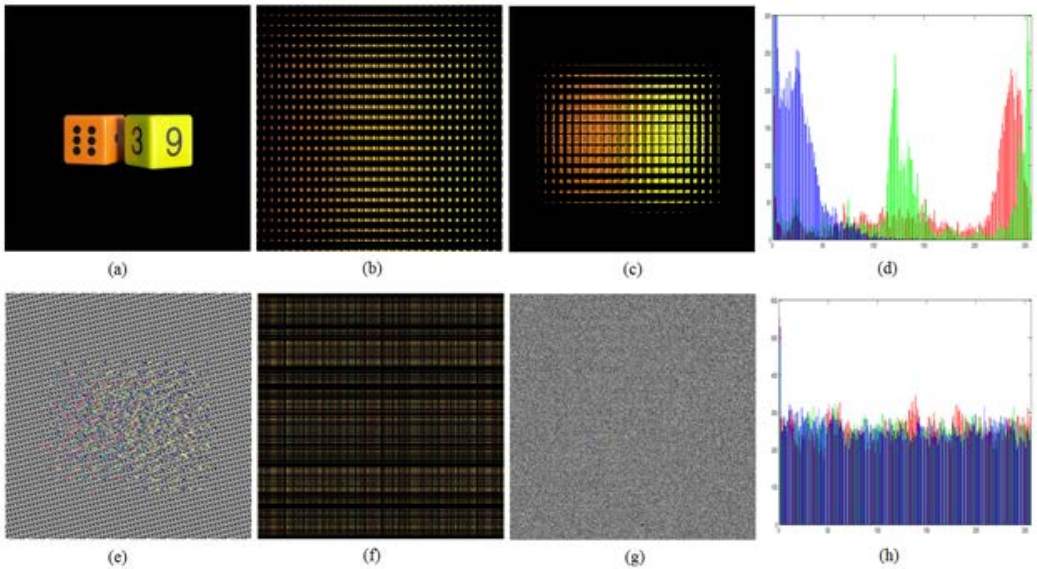}
\caption{Simulation results on plain-image ``Dice": (a) the plain-image ``dice", (b) the direct picked-up EIA, (c) the depth-converted EIA, (d) the histogram of the plain-image, (e) the encrypted image by the pseudo-random mask, (f) the encrypted image by chaos map with initial parameters ($x_0 = 0.1775727$ and
$\rho_0=3.5725212$), (g) the encrypted image by the proposed scheme, (h) the histogram of cipher-image of the proposed scheme.}
\label{fig:EIA_Diagram1}
\end{figure}

To show the effectiveness of the proposed encryption scheme, we performed a series of experiments. The computer pickup structure is shown in Fig.~\ref{fig:EIA_Diagram}(b). The pinhole array used in this experiment is composed of $30\times 30$ pinholes, the interval between pinholes is 1.08 mm and the gap $g$ between the EIA and the pinhole array is 3 mm. Several typical images, such as gray and color images are used as the test images. Each of the test images has $900\times 900$ pixels. The pickup distance between the pinhole array and the input image is set as $l= 69$ mm. In the encryption structure shown in Fig.~\ref{fig:flowchart}, we first synthesize the EIA by the computational pickup based on the ray geometry. The resultant EIA is shown in Fig.~\ref{fig:EIA_Diagram1}(b). Then, the directly picked-up EIA is transformed by using the proposed smart mapping method in order to obtain a depth-converted EIA, which is shown in Fig.~\ref{fig:EIA_Diagram1}(c). To demonstrate the effectiveness of the image encryption scheme, we comparatively analyze the performance of the proposed image encryption result shown in Fig.~\ref{fig:EIA_Diagram1}(g) with the pseudo-random mask results shown in Fig.~\ref{fig:EIA_Diagram1}(e) and the chaotic logistic maps encryption result shown in Fig.~\ref{fig:EIA_Diagram1}(f). From the simulation results, it is clear that the proposed encoding method that combining the use of pseudo-random mask and chaotic logistic maps can provide a high quality encrypted image.

\subsection{Histogram analysis}

The cipher-image histogram analysis is one of the most straight forward methods for illustrating the image encryption quality. Since a practical image encoding method tends to encode a plain-image to random-like noise, it is desired to show a uniformly distributed histogram for the cipher-image. Fig.~\ref{fig:EIA_Diagram1}(d) and (h) display the histograms of the original and encrypted images. The histogram of the encrypted image is uniformly distributed. In other words, the cipher-image is random-like. Fig. 6 shows the simulation results using a test image ``Car".
Fig. 7 shows several ciphertext histograms from some kinds of the encrypted images. It is worthwhile to note that the images cover the format from gray and RGB color images. From the simulation results, it is clear that although the histograms of the plaintext images are very steep, the histograms of the encrypted images become very flat after encryption. After a series of parallel experiments, we conclude that the histograms of the original images are dramatically different from the ones after encryption. It is difficult for an attacker to obtain a useful clue from the statistical characteristic. Thereby it can bring us considerable capacity of resisting statistical attack.
\begin{figure}
\center
\includegraphics[width=1.1\imagewidth]{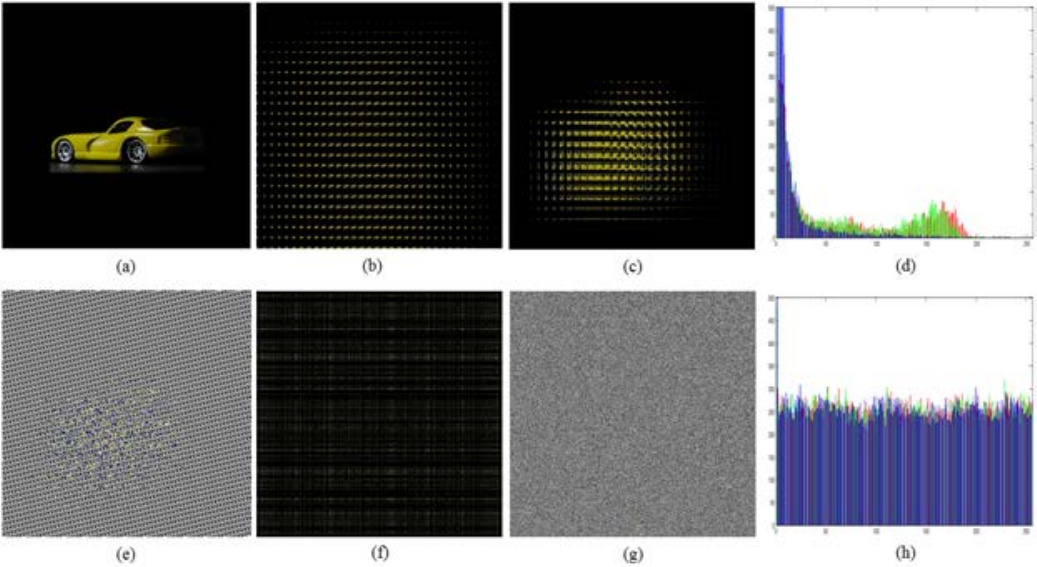}
\caption{Experiment results of a sample plain-image: (a) the plain-image ``Car"; (b) the direct picked-up EIA; (c) the depth-converted EIA; (d) the histogram of the plain-image; (e) the encrypted image by the pseudo-random mask; (f) the encrypted image by chaotic maps with initial parameters, $x_0 = 0.1775727$ and $\rho_0=3.5725212$; (g) the encrypted image with the proposed scheme; (h) the histogram of the proposed encrypted image.}
\label{fig:expHistogram}
\end{figure}

\begin{figure}
\center
\includegraphics[width=\imagewidth]{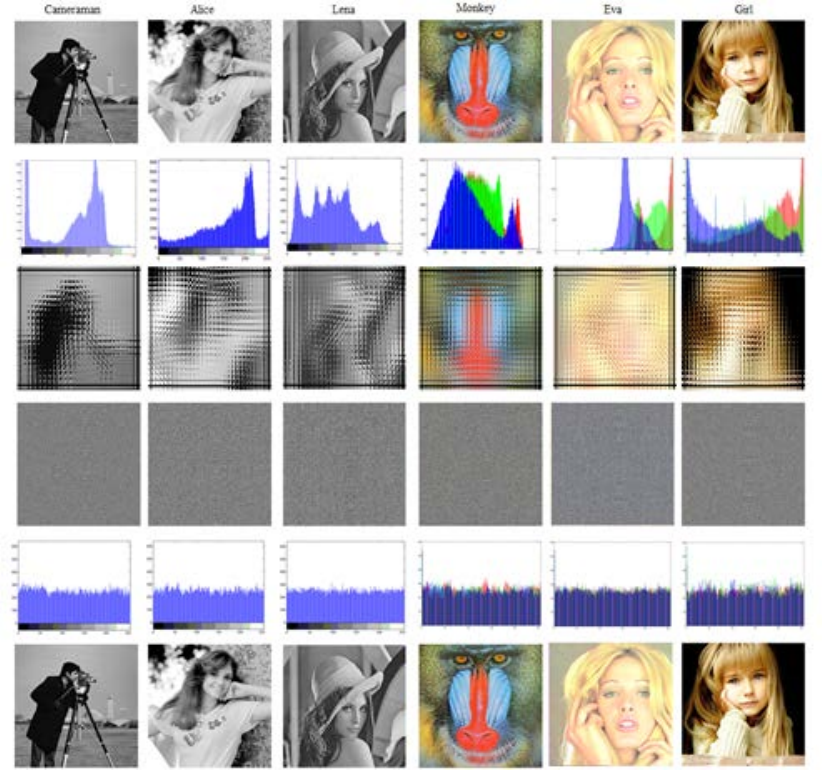}
\caption{Histograms analysis on the encrypted images.}
\label{fig:expHistogram}
\end{figure}

\subsection{Key space and sensitivity analysis}

\begin{figure}
\center
\includegraphics[width=\imagewidth]{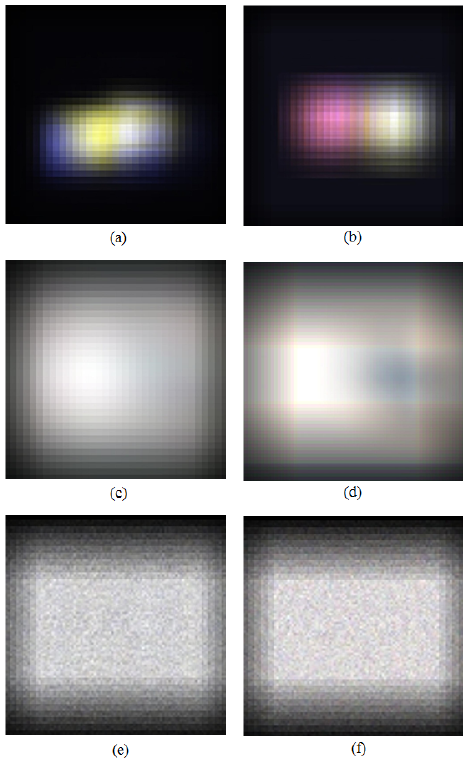}
\caption{Reconstructed images using partially incorrect keys: (a) the reconstructed images corresponding to plain-image ``Car" with the incorrect pickup distances $l=30\neq 69 mm$; b) the reconstructed images corresponding to plain-image ``Dice" with the incorrect pickup distances $l=30\neq 69 mm$;
(c) the reconstructed images corresponding to plain-image ``Car" with incorrect key, $x_0 = 0.1675727\neq 0.1775727$; (d) the reconstructed images corresponding to plain-image ``Dice" with incorrect key, $x_0 = 0.1675727\neq 0.1775727$; (e) the reconstructed image corresponding to plain-image ``Car" with incorrect rule $R=(90 90 150 90 90 90 150 90)$; (f) the reconstructed image corresponding to plain-image ``Dice" with incorrect rule $R=(90 90 150 90 90 90 150 90)$.}
\label{fig:KeySensitive}
\end{figure}

The proposed encryption scheme provides a high degree of security because of the multiple security key space: initial parameters of CGII pickup system, pickup distance $l$, the gap $g$; the input initial parameters of chaotic logistic maps $(x_0, \rho)$; the pseudo-random mask generation CA rule $\mathrm{R}=(150 90 150 90 90 90 150 90)$. To analyze a case where partial keys are obtained by attackers, decryption tests are performed as shown in Fig.~\ref{fig:KeySensitive}. Fig.~\ref{fig:KeySensitive}(a) and (b) show the reconstructed images ``Car" and ``Dice" with the incorrect pickup distances  mm, respectively. Fig.~\ref{fig:KeySensitive}(c) and (d) show the reconstructed images using the incorrect initial parameter ($x_0 = 0.1675727$). Figure~\ref{fig:KeySensitive}(e) and (f) show the reconstructed images using an incorrect CA, $\mathrm{R}=(90 90 150 90 90 90 150 90)$.
From the simulations results displayed on Fig.~\ref{fig:KeySensitive}, it is clear that there is a significant difference between the reconstructed and the original images, although when attackers have obtained the keys of the two components and only one key is unknown. These results clearly show that the proposed encoding scheme is very sensitive to the encryption keys.

\subsection{Peak signal-to-noise ratio analysis}

\begin{figure}
\center
\includegraphics[width=\imagewidth]{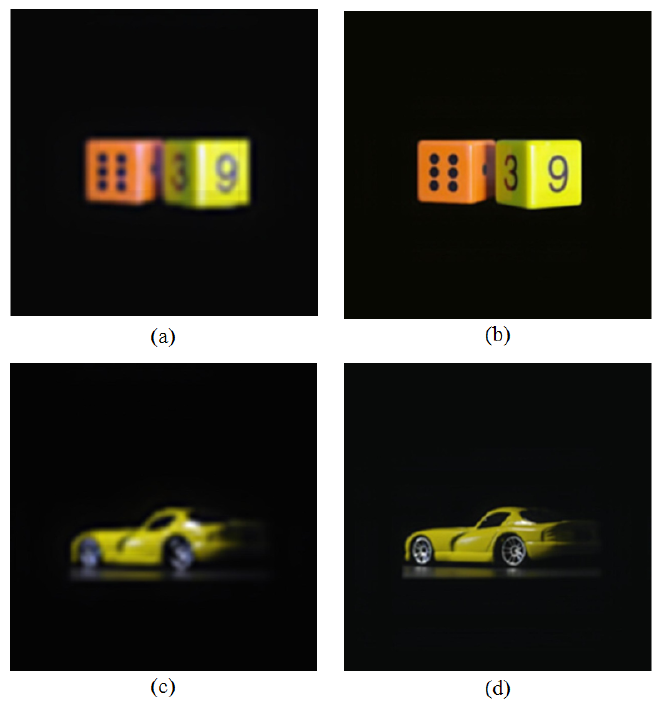}
\caption{Reconstructed images with all correct keys: (a) the reconstructed image corresponding to plain-image ``Dice" using the direct picked-up EIA, $z = l = 69$ mm; b) the reconstructed image corresponding to plain-image ``Dice" using the direct picked-up EIA, $z = l = 69$ mm;
(c) the reconstructed image corresponding to plain-image ``Car" with incorrect key, $x_0 = 0.1675727\neq 0.1775727$; (d) the reconstructed image corresponding to plain-image ``Car" using the depth-converted EIA, $z = d-l = 90-69 = 21$ mm.}
\label{fig:PSNRanalysis}
\end{figure}

To verify the decryption quality of the proposed scheme, the peak signal-to-noise ratio (PSNR) is introduced to provide completed of information of the pixels distribution respect to the original image. The PSNR with high values correspond to strong similarity between the encrypted images and the original ones. The PSNR is defined as
\begin{equation*}
PSNR=10\log_{10}\left(\frac{Max^2}{MSE}\right),
\end{equation*}
\begin{equation*}
MSE=\frac{1}{MN}\sum_{i=0}^{M-1}\sum_{j=0}^{N-1}(O(x, y)-O_r(x, y))^2,
\end{equation*}
where $Max$ denotes the is the maximum possible pixel value of the image, $O$ and $Or$ represent the original and the encrypted images, respectively.
Figure~\ref{fig:PSNRanalysis}(a) and (c) show the reconstructed images based on the conventional CGII system. Figure~\ref{fig:PSNRanalysis}(b) and (d) show the reconstructed images based on the proposed scheme. By visually comparing these two methods, it is clear that our proposed method provides better resolution. The PSNR values for each channel of the reconstructed images ``Dice" and ``Car" with the conventional CGII method are 27.2645, 25.2198, 27.9980 dB, and 26.3327, 25.0219, 26.0017 dB, respectively. However, the PSNR values of ``Dice" and ``Car" with the proposed method are 38.4418, 36.1249, 39.3369 dB, and 38.0120, 37.4426, 39.0579 dB, respectively. From the calculated PSNR values, it can be seen that PSNR values of image ``Dice" and ``Car" with the proposed scheme outperforms the conventional CGII by $43.7964\%$ and $48.0326\%$, respectively.

To further demonstrate the effectiveness of the proposed algorithm, we comparatively analyzed the proposed PSNR values with \cite{Lixw:integral:Optik2014} and \cite{Piao:integral:OLE2009} and the results of PSNR are recorded in Table~\ref{table:PSNR}. In comparison with the simulation results, it is noticeable that the proposed method has very excellent PSNR scores. It outperforms other recent integral imaging-based image encoding methods.

\begin{table}[ht]
\centering
\caption{PPSNR values for the decrypted images.}
\begin{tabular}{l| c| c| c| l}\hline
Proposed &	Images &    PSNR   &  	PSNR   & PSNR\\\hline
Gray&	``Cameraman" &	       &	38.5413&	 \\
    &	``Alice"     &         &	37.0023&	 \\
    &	``Lena"      &	       &	38.1654&	 \\\hline
Color&	        & R            &	G     &	  B    \\
&	``Dice"   &	38.4418 &	36.1249&	39.3369\\
&	``Car"    & 38.0120 &	37.4426&	39.0579\\
&	``Monkey" &	39.0013 &	38.9120&	38.7715\\
&	``Eva"    &	37.5698 &	36.0013&	27.3395\\
&	``Girl"   & 38.4432 &	37.2895&	37.6984\\\hline
Ref. \cite{Lixw:integral:Optik2014}    \\\hline
Gray&	``Cameraman"&	&	26.3387&	\\
    &	``Alice"    &   &	25.9874&	\\
    &	``Lena"     &	&	25.3391&	\\\hline
Color&	     & R       &	G      &	  B    \\
&	``Dice"  &	27.6653&	26.9845&	27.3695\\
&	``Car"   &  28.6541&	27.3290&	29.3381\\
&	``Monkey"&	26.5589&	24.6874&	25.9842\\
&	``Eva"   &	29.5401&	28.6245&	28.3025\\
&	``Girl"  & 27.3394&	26.3365&	28.6541\\\hline
Ref. \cite{Piao:integral:OLE2009}	           \\\hline
Gray &	``Cameraman" &	&	22.2184&	\\
     &	``Alice"     & &	24.6891&	\\
     &	``Lena"      &	&	23.6410&	\\\hline
Color&	        & R        &	G     &	  B    \\
&	``Dice"  &	20.1148&	19.8469&	21.3368\\
&	``Car"   & 22.5984&	21.0025&	23.6584\\
&	``Monkey"&	21.5598&	20.6589&	23.6640\\
&	``Eva"   &	23.6694&	22.6513&	24.6635\\
&	``Girl"  & 22.6513&	21.6635&	23.9841\\\hline
\end{tabular}
\label{table:PSNR}
\end{table}

\subsection{Resistance to data loss attack}

\begin{figure}
\center
\includegraphics[width=\imagewidth]{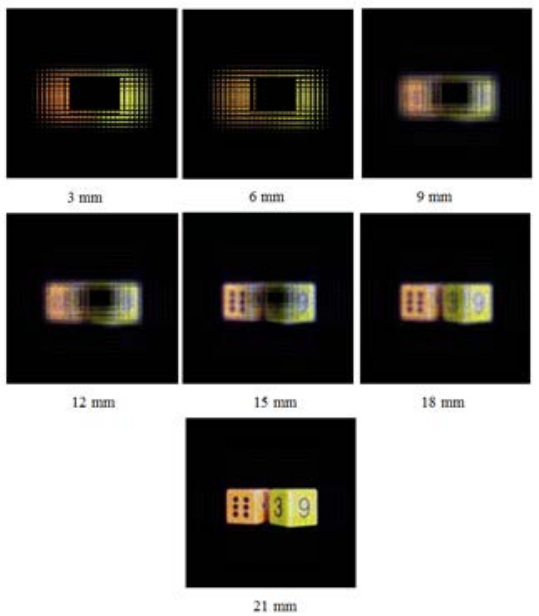}
\caption{Reconstructed images from an attacked EIA, where intervals are specified along the output plane with interval of 3mm.}
\label{fig:DataSensitive}
\end{figure}

\begin{figure}
\center
\includegraphics[width=\imagewidth]{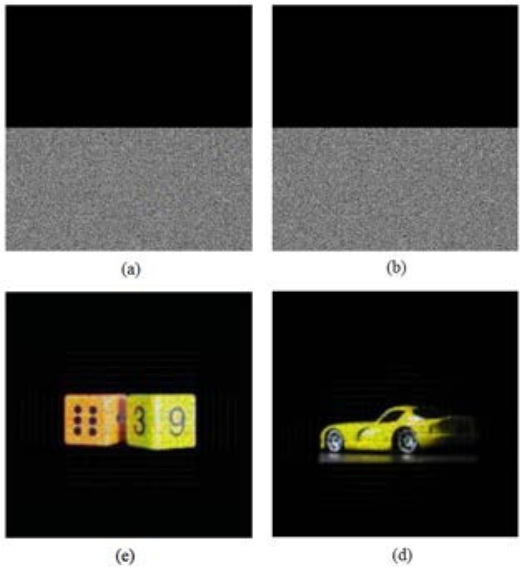}
\caption{Robustness tests against occlusion attack: (a) the attacked version of the cipher-image corresponding to image ``Dice" with $D= 50\%$;
(b) the attacked version of the cipher-image corresponding to image ``Car" with $D= 50\%$;
(c) the reconstructed plain-image from (a); (d) the reconstructed plain-image from (b).}
\label{fig:Reconstruct}
\end{figure}

We also perform the computer simulation to check the tolerance of the proposed method to data loss attack. The proposed scheme provides high robustness because of the hologram-like property of EIA. The EIA is composed of many elemental images and each of them nearly possesses the full attribute of the input image. Although most data of EIA has been destroyed, the image can be successfully reconstructed from the remaining information of EIA. Figure \ref{fig:DataSensitive} shows a sample result of the image reconstruction process with the data loss attack, in which about $30\%$ pixels of the EIA are occluded. From the results displayed in Fig.~\ref{fig:DataSensitive}, it is clear that the image can be successfully reconstructed, even though most information of the elemental images is occluded. Figure~\ref{fig:Reconstruct} shows the occluded encrypted images and the corresponding reconstructed images, in which $50\%$ of the encrypted image are set to be occluded. The EIA provides robustness to the distortions due to property of data redundancy of the elemental images. It is useful when transmit digital images through network, especially for only loading part of the encoded images in the case of heavy traffic.
The PSNR values of the reconstructed images under occlusion attack are recorded in Table~\ref{table:occlusionPSNR}. In comparison results of the occlusion attack, it is clear that the PSNR values are superior to the CGII methods \cite{Lixw:integral:Optik2014,Piao:integral:OLE2009}. The property of data redundancy of the elemental images improves the robustness of the encrypted image. Meanwhile, the depth-converted EIA of our scheme provides high image resolution.

\begin{table}[ht]
\centering
\caption{PSNR for decrypted images under the occlusion attack.}
\begin{tabular}{l| c| c| c| l}\hline
$¡®v¡¯$&	Images &  R	      &  G	   &  B      \\\hline
10\%&	``Dice"&	35.3395&	34.2690&	36.2678\\
10\%&	``Car" & 36.8873&	35.3369&	37.6231\\
20\%&	``Dice"&	31.2159&	31.8879&	32.0145\\
20\%&	``Car" & 32.5694&	30.1591&	33.6620\\
50\%&	``Dice"&	28.3361&	27.1562&	28.3210\\
50\%&	``Car" & 27.3361&	26.3845&	28.6231\\
70\%&	``Dice"&	21.4451&	20.2894&	22.0101\\
70\%&	``Car" & 20.1548&	19.3258&	22.6541\\\hline
Ref. \cite{Lixw:integral:Optik2014}    \\\hline
10\%&	``Dice"&	24.6638&	23.6512&	24.6513\\
10\%&	``Car" & 23.3367&	23.6584&	24.1121\\
20\%&	``Dice"&	18.3247&	17.6784&	18.8842\\
20\%&	``Car" & 18.9874&	16.3210&	18.3328\\
50\%&	``Dice"&	14.6584&	13.5985&	14.3362\\
50\%&	``Car" & 13.2012&	12.0395&	14.3652\\
70\%&	``Dice"&	11.3541&	10.6542&	10.8879\\
70\%&	``Car" & 10.3218&	9.9951 &    11.3217\\\hline
Ref. \cite{Piao:integral:OLE2009}	           \\\hline
10\%&	``Dice"&	21.3395& 20.3984&	22.3675\\
10\%&	``Car" & 21.6354& 19.3265&	21.0020\\
20\%&	``Dice"&	16.5214& 14.3215&	15.9841\\
20\%&	``Car" & 15.3510& 14.3028&	16.9842\\
50\%&	``Dice"&	12.6657& 12.6984&	13.5547\\
50\%&	``Car" & 12.6540& 12.5462&	13.9982\\
70\%&	``Dice"&	8.7745 & 9.3236	&    10.8891\\
70\%&	``Car" & 9.3984 & 8.9902	&    10.0017\\\hline
\end{tabular}
\label{table:occlusionPSNR}
\end{table}

\subsection{Noise attack analysis}

\begin{figure}
\center
\includegraphics[width=\imagewidth]{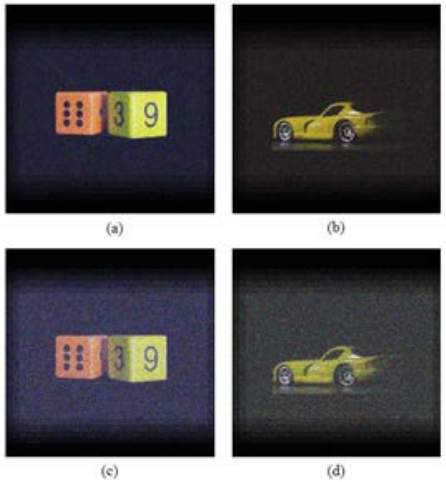}
\caption{Robustness test against Gaussian noise attack: (a) the reconstructed plain-image ¡®Dice¡¯ with $v = 0.1$;
(b) the reconstructed plain-image ¡®Car¡¯ with $v = 0.1$; (c) the reconstructed plain-image ¡®Dice¡¯ with $v = 0.5$;
(d) the reconstructed plain-image ¡®Car¡¯ with $v = 0.5$.}
\label{fig:Noise}
\end{figure}

The simulations are performed to test the robustness of the proposed encryption scheme against additive noise attack, where Gaussian noise of a zero mean
was adopted. For Gaussian noise attack, we assume that the noise is added into the encrypted image using equation
\begin{equation*}
O'_e=O_e(1+vN),
\end{equation*}
where $O_e$ and $O'_e$ represent the encrypted image and the noise affected encrypted image, respectively. Varible $v$ denotes the noise strength and $N$ is Gaussian noise with zero-mean. Figure~\ref{fig:Noise} shows the reconstructed images when $v$ equals to 0.1 and 0.5. From simulation results displayed in Fig.~\ref{fig:Noise}, the decrypted images can be recognized despite of the noise interference. Although when $v = 0.5$, the reconstructed images can still be clearly recognized. The calculated PSNR values are tabulated in Table~\ref{table:GaussianPSNR}. The PSNR values of the images generated by the proposed scheme are obviously higher than that of the conventional methods under Gaussian noise attack. From the simulation results, we can conclude that the robustness of our scheme is superior to the conventional schemes.

\begin{table}[ht]
\centering
\caption{PSNR values of decrypted images under Gaussian noise attack.}
\begin{tabular}{l| c| c| c| l}\hline
$¡®v¡¯$&	Images  &  R	      &  G	    &  B      \\\hline
0.1&	``Dice" &  32.8861 & 31.0282    &  33.2589\\
0.1& 	``Car"  &  33.6491 &	32.5984 &  33.9658\\
0.5&	``Dice" &  27.3364 & 26.3218	&  28.3361\\
0.5&	``Car"  &  26.3201 & 25.0021	&  27.4456\\
0.8&	``Dice" &  21.3384 & 20.1984	&  22.3651\\
0.8&	``Car"  &  20.1845 & 19.2314	&  21.3654\\
\hline
Ref. \cite{Lixw:integral:Optik2014}    \\\hline
0.1&	``Dice" &	26.3124&	25.3310&	25.3981\\
0.1&	``Car"  &   25.6987&	24.6518&	25.3214\\
0.5&	``Dice" & 	22.3846&	20.1694&	22.3987\\
0.5&	``Car"  &   21.3654&	20.9842&	22.3694\\
0.8&	``Dice" &	17.3994&	16.9984&	16.3217\\
0.8&	``Car"  &   18.6541&	17.6945&	17.6980\\\hline
Ref.~\cite{Piao:integral:OLE2009}    \\\hline
0.1&	``Dice" &	24.3695&	23.6941&	25.4639\\
0.1&	``Car"  & 24.3641&	24.1654&	24.6999\\
0.5&	``Dice" &	19.3657&	18.3654&	19.3698\\
0.5&	``Car"  & 18.6945&	18.2011&	19.0298\\
0.8&	``Dice" &	14.6891&	13.3984&	14.9541\\
0.8&	``Car"  & 15.3349&	14.6469&	15.5943\\\hline
\end{tabular}
\label{table:GaussianPSNR}
\end{table}

\section{Conclusion}

In this paper, an image encryption scheme was proposed by employing smart mapping in the CGII system. We explained the pixel-interference problem due to large magnification in the superposition process of CIIR. To overcome the pixel-interference problem, the magnification factor is required to be minimized. Based on this, we proposed a smart mapping based depth-conversion scheme to lessen the reconstruction distance and consequently reduce the magnification factor. Through experimental results and security analysis, we demonstrated that our proposed scheme possesses a better encryption, a larger key space and is highly sensitive to the secret key. Furthermore, the proposed scheme can also provide super resolution and high robustness resist attacks.

\section*{Acknowledgement}

This research is supported by Ministry of Culture, Sports and Tourism (MCST) and Korea Creative Content Agency (KOCCA) in the Culture Technology (CT) Research \& Development Program 2014.
C. Li was supported by the Distinguished Young Scholar Program of Hunan Provincial Natural Science Foundation of China (No.~2015JJ1013).

\bibliographystyle{IEEEtran}
\bibliography{EXW}
\end{document}